\documentclass[twoside]{article}
\usepackage{fleqn,espcrc2}
\usepackage{epsfig}
\usepackage{graphicx}
\usepackage[figuresright]{rotating} 
\title{Atmospheric neutrino oscillations from 
upward throughgoing muon multiple scattering in MACRO}
\author{
\begin{center}
{\bf The MACRO Collaboration  } \\
\nobreak\bigskip\nobreak
M.~Ambrosio$^{12}$, 
R.~Antolini$^{7}$, 
D.~Bakari$^{2,17}$,
A.~Baldini$^{13}$, 
G.~C.~Barbarino$^{12}$, 
B.~C.~Barish$^{4}$, 
G.~Battistoni$^{6,a}$,
Y.~Becherini$^{2}$,
R.~Bellotti$^{1}$, 
C.~Bemporad$^{13}$
% ...
\\
\vspace{.5 cm}
\footnotesize
1. Dipartimento di Fisica dell'Universit\`a  di Bari and INFN, 70126 Bari, Italy \\
2. Dipartimento di Fisica dell'Universit\`a  di Bologna and INFN, 40126 Bologna, Italy \\
3. Physics Department, Boston University, Boston, MA 02215, USA \\
4. California Institute of Technology, Pasadena, CA 91125, USA \\
5. Department of Physics, Drexel University, Philadelphia, PA 19104, USA \\
6. Laboratori Nazionali di Frascati dell'INFN, 00044 Frascati (Roma), Italy \\
7. Laboratori Nazionali del Gran Sasso dell'INFN, 67010 Assergi (L'Aquila), Italy \\
8. Depts. of Physics and of Astronomy, Indiana University, Bloomington, IN 47405, USA \\
9. Dipartimento di Fisica dell'Universit\`a  dell'Aquila and INFN, 67100 L'Aquila, Italy\\
10. Dipartimento di Fisica dell'Universit\`a  di Lecce and INFN, 73100 Lecce, Italy \\
11. Department of Physics, University of Michigan, Ann Arbor, MI 48109, USA \\
12. Dipartimento di Fisica dell'Universit\`a  di Napoli and INFN, 80125 Napoli, Italy \\
13. Dipartimento di Fisica dell'Universit\`a  di Pisa and INFN, 56010 Pisa, Italy \\
14. Dipartimento di Fisica dell'Universit\`a  di Roma "La Sapienza" and INFN, 00185 Roma, Italy \\
15. Physics Department, Texas A\&M University, College Station, TX 77843, USA \\
16. Dipartimento di Fisica Sperimentale dell'Universit\`a  di Torino and INFN, 10125 Torino, Italy \\
17. L.P.T.P, Faculty of Sciences, University Mohamed I, B.P. 524 Oujda, Morocco \\
18. Dipartimento di Fisica dell'Universit\`a  di Roma Tre and INFN Sezione Roma Tre, 00146 Roma, Italy \\
\end{center}
}
\begin{document}
\begin{abstract}
{The energy of atmospheric neutrinos detected by MACRO was estimated
using multiple coulomb scattering of upward throughgoing muons.
This analysis allows a test of atmospheric neutrino oscillations,
relying on the distortion of the muon energy distribution.} 
These results have been combined with those coming from the upward throughgoing muon
angular distribution only. Both analyses are independent of the neutrino
flux normalization and
provide strong evidence, above the 4$\sigma$ level, in favour of neutrino
oscillations.\\
PACS: 14.60Lm,14.60.Pq, 25.30.Mr
\end{abstract}
\maketitle
\section{Introduction}
\footnotetext{
$a$ Also INFN Milano, 20133 Milano, Italy \\
$b$ Also IASF/CNR, Sez. di Bologna, Italy \\
$c$ Also Department of Physics, Pennsylvania State University, University
Park, PA 16801, USA \\
$d$ Also Universit\`a  di Trieste and INFN, 34100 Trieste, Italy \\
$e$ Also U. Minn. Duluth Physics Dept., Duluth, MN 55812 \\
$f$ Also Dept. of Physics, MIT, Cambridge, MA 02139 \\
$g$ Also Intervideo Inc., Torrance CA 90505 USA \\
$h$ Also Department of Physics, SLIET,Longowal,India\\
$i$Also Dipartimento di Fisica dell'Universit\`a  della Calabria, Rende 
(Cosenza), Italy \\
$l$ Also RPD, PINSTECH, P.O. Nilore, Islamabad, Pakistan \\
$m$ Also Institute for Space Sciences, 76900 Bucharest, Romania \\
$n$ Also Universit\`a  della Basilicata, 85100 Potenza, Italy \\
$o$ Also Dip. di Energetica, Universit`a di Roma, 00185 Roma, Italy.\\
$p$ Also Dipartimento di Scienze e Tecnologie Avanzate, Universit\`a  del 
Piemonte Orientale, Alessandria, Italy \\
$q$ Also Resonance Photonics, Markham, Ontario, Canada\\
$*$ Corresponding authors: SCAPPARONE@BO.INFN.IT,\\SIOLI@BO.INFN.IT}

The results obtained by experiments looking for neutrinos
coming from natural or artificial far sources gave 
the first indication of physics beyond the standard model. Neutrino flavor
changing is in fact the most straightforward explanation for 
electron and muon neutrino disappearance and for the 
evidence of non electron neutrinos from the sun.  
In this scenario, the study of atmospheric neutrinos plays an important
role to improve our understanding of
the neutrino oscillation mechanism.

The MACRO detector studied~\cite{macroneutrino,lownu,sterile} three categories of events as
shown in Fig. 1: (1)
upward throughgoing muons, (2) upward semicontained muons, and (3+4) upward stopping
muons plus downward semicontained muons. The (3+4) categories cannot be
separated experimentally, because the events have only one time
measurement. However, upward stopping muons and downward semicontained
muons have similar parent neutrino energy.
\begin{figure}[t]
 \begin{center}
  \vskip 3.1cm
  \mbox{\epsfig{file=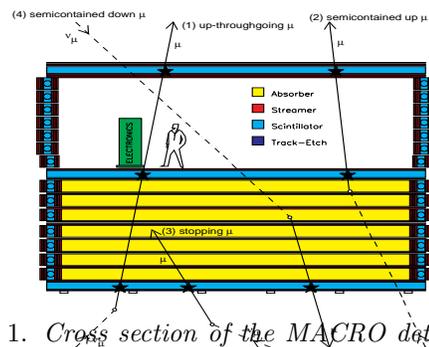,width=7.5cm,height=5.5cm}}
  \vskip -3.1cm
  \caption{\it Cross section of the MACRO detector and topologies of
events induced by neutrino interactions.\label{fig:figuramacro}}
\vspace{-0.7cm}
 \end{center}
\end{figure}

An atmospheric muon neutrino deficit was found in all of the three
categories and 
an angular distribution distortion was found in category
(1). Such results   
are explained by the neutrino oscillation hypothesis with parameters 
$\Delta {\rm m^2} = 2.5 \times10^{-3} {\rm eV^2}$ and 
$\sin^22\theta$=1, in good agreement with the Super-Kamiokande 
results~\cite{SuperKneu}.
A detailed study of the upward throughgoing muon angular distribution
by MACRO~\cite{sterile} and by Super-Kamiokande~\cite{SK}
allowed the exclusion at the 99$\%$C.L. of the 
muon neutrino into sterile
neutrino oscillation mechanism, compared to the muon neutrino into the
tau neutrino. 

Considering a two flavor neutrino oscillation, the probability for a 
$\nu_{\mu}$ to oscillate to $\nu_{\tau}$ is given by:  
\begin{equation}
P_{\nu_{\mu}\rightarrow\nu_{\tau}}=\sin^{2}2\theta \sin^{2}
\left(\frac{1.27L_{\nu}\Delta
{\rm m^{2}}}{E_{\nu}}\right)
\end{equation}
where $L_{\nu}$(km) is the 
distance between the neutrino production and interaction points and
$E_{\nu}$($\rm GeV$) is the
neutrino energy. 
The different
categories of events quoted above have median energy
from  $4\, \hbox{\rm GeV}$ to $50\, \hbox{\rm GeV}$, which
provides evidence of the dependence on  
neutrino energies, as required by the
oscillation hypothesis. 
In this paper we address for the first time the estimate of the upward throughgoing muon
energy by using multiple scattering. 
The implementation of this method to
the MACRO data is discussed in Section 2. Two different analysis were
performed. The first one, using the tracking system in digital mode, is
described in Section 3.
Section 4 shows the results obtained with the streamer tube in drift mode,
where electronic readout of the hit time was used to improve the position
resolution 
and thereby the range of muon energies that could be estimated. 

\section{The muon energy estimate}
The MACRO detector was extensively described in~\cite{macrotecnico}. 
The detector consisted of a lower half~(the {\it lower detector}) and
an upper half~(the {\it ``Attico''}). The {\it lower detector} had
10 horizontal planes of streamer tubes separated by either 
crushed rock absorbers ($60\, \hbox{\rm g/${\rm cm^2}$}$) or (at its top and bottom) a plane of scintillator 
counters. The $Attico$ had  
4 horizontal planes of streamer tubes separated into two groups
with a plane of scintillator counters between them.
 
Upward throughgoing muons are mainly
produced in neutrino deep inelastic scattering (DIS) in the rock below the
detector. 
The recoil hadrons are
lost and the muon energy is degraded in the propagation
to the detector. Nevertheless, Monte Carlo simulations
show a linear relation between the parent neutrino
energy $E_{\nu}$ and the muon residual energy $E_{\mu}$ at detector level.

The momentum resolution obtainable from multiple coulomb scattering\ (MCS) 
measurements is the
result of two different contributions: the number of sampling planes $N$
and the space resolution of the detector $\sigma_{0}$.
In MACRO the number of tracking planes interleaved with
rock absorbers gives $N=8$. The resolution
of the streamer tube system used in digital mode 
is $\sigma_{0}\simeq$$1\, \hbox{\rm cm}$
and was improved to $\sigma_{0}\simeq$$0.3\, \hbox{\rm cm}$ by using the drift time.
The other six horizontal tracking planes are separated by a negligible
amount of material, and do not contribute to the MCS measurements.

In a tracking detector with equispaced tracking planes, separated by 
a given absorber and neglecting the energy losses, 
the characteristic momentum scale can be written as~\cite{librosioli}:\\
\begin{equation}
  p_{MCS}(\rm GeV)=\frac{0.015\Delta\sqrt{\Delta/X_{0}}}{\sigma_{0}}
\end{equation}
where $\Delta$ is the distance between tracking planes and $X_{0}$
is the material radiation length.
For p$<$$p_{MCS}$, the main limitation to the momentum reconstruction
comes from the number of sampling planes while for p$>$$p_{MCS}$ the space
resolution dominates.
Under the conditions quoted above the 
relative momentum error~\cite{librosioli} can be written as:
\begin{equation}
  \rm\frac{\sigma_{p}}{p}\rightarrow\frac{1}{\sqrt{2N}}
\end{equation}
for p$<<$$p_{MCS}$, and
\begin{equation}
  \rm\frac{\sigma_{p}}{p}\rightarrow\frac{1}{\sqrt{2N}}\left(\frac{p}{p_{MCS}}\right)^2,
\end{equation}
for p$>>$$p_{MCS}$. 
As shown in \cite{librosioli} the 1/p distribution
approaches a Gaussian only for N$\geq$30; therefore the 
momentum error for MACRO is not expected to have a Gaussian behaviour.
The eight horizontal 
streamer tube planes interleaved with rock 
absorbers (planes 2 to 9, from the bottom)
are equispaced and have the same space resolution, hence
$p_{MCS}^{MACRO}$=$2.2\, \hbox{\rm GeV/c}$.
Since about 90$\%$ of upward throughgoing muons 
have p~$>$~$2.2\, \hbox{\rm GeV/c}$, 
the intrinsic chamber resolution dominates.
The 
MCS-based momentum estimates scale therefore as the square of the space 
resolution $\sigma_{o}$.

The r.m.s.
of the lateral displacement of a relativistic muon crossing a layer of material
with depth $X$ 
is proportional to the inverse of muon momentum $p_{\mu}$\cite{DPB}:
\begin{equation}
  \sigma_{proj}^{MCS} \simeq \frac{X}{\sqrt{3}}\frac{13.6\, \hbox{\rm MeV}} {p_{\mu}\beta c}
  \sqrt{\frac{X}{X_{o}}}\left(1+0.038\ln\frac{X}{X_{o}}\right).
\end{equation}
In MACRO, X$\simeq$25$X_o$/$\cos\Theta$, giving on the vertical
$\sigma_{proj}^{MCS}\simeq$$10\, \hbox{\rm cm}$/$E_{\mu}(\rm GeV)$. Therefore, a saturation point above
$10\, \hbox{\rm GeV}$ requires a space resolution better than $1\, \hbox{\rm cm}$.
An improvement in the space resolution increases the maximum energy
value (saturation point) above which the MCS method is not effective.

Consequently we decided to 
improve the streamer tube resolution:
in the second analysis, described in Section 4, we 
took advantage
of the QTP-TDC electronics, developed for magnetic monopole
searches~\cite{macrotecnico}. 
In this case, we obtained an improved space resolution, 
$\sigma_{x}$$\sim$$0.3\, \hbox{\rm cm}$\cite{macronim}, which allowed us to estimate 
the muon energy up to $E_{\mu}$$\simeq$$40\, \hbox{\rm GeV}$.

For both analyses, we used the whole sample of upward throughgoing muon events
collected with the complete apparatus in a period of data taking 
equivalent to 5.5  years of live time. We studied upward muon 
events selected by the time-of-flight measured by planes of
scintillators combined with the standard MACRO tracking algorithm.
The original upward throughgoing muon data set has been described 
previously~\cite{sterile}.
To make a comparison between real data and
expectation we performed a Monte Carlo simulation using the Bartol 
neutrino flux~\cite{Bartol} and the GRV94 DIS parton distributions
~\cite{GRV94} for deep inelastic scattering. For low energy 
neutrino interactions we used the
cross sections given in~\cite{lipari}.
The energy loss for muons propagating through rock is taken from 
Lohmann et al.~\cite{lohman} 
while the muon simulation inside the detector was performed with GMACRO 
(the GEANT 3.21 based detector simulation). 
For the second
analysis, where the streamer system is used in drift mode, another
simulation chain has been implemented. In this case neutrino
interactions are randomly distributed in a rock semi-sphere below the
detector.
Muons are then transported to the detector
using the FLUKA99 package\cite{fluka}. 
A Monte Carlo statistics corresponding to a live time of $\simeq$2700 years 
was produced~(500 equivalent experiments).
This simulation was compared with that used in\cite{macroneutrino,sterile},
obtaining a satisfactory result.
\section{The first analysis: streamer tubes in digital mode}
Because of MCS in the {\it lower detector}, 
with this analysis we expect a measurable 
deflection between the incoming and the outgoing track directions 
for muons with energies smaller than $\sim$$10\, \hbox{\rm GeV}$. 
This corresponds to an effective 
arm-lever of $\sim$$4\, \hbox{\rm m}$ between the {\it lower detector} and 
the $Attico$\cite{macrotecnico}.

The eight lowest streamer tube planes were used, through a track refit, 
to estimate the direction of the incoming muon. 
The five upper streamer tube planes in the {\it lower detector}
and the four $Attico$ planes were used to estimate the direction of the outgoing muon.
The distance $r_w$ between the intercepts of the two tracks in 
the $z=0$ plane, and the difference $\Delta \Phi$ between 
the two slopes depend on the muon energy  $E_\mu$.
We divided the upward throughgoing muons in three sub samples, 
according to the values of $r_w$ and $\Delta \Phi$: 
sample {\it L= Low energy}, if $r_w>$$3\, \hbox{\rm cm}$ and $\Delta \Phi> 0.3^\circ$; 
{\it H= High energy} if  $r_w \le $$3\, \hbox{\rm cm}$ and $\Delta \Phi \le 0.3^\circ$.
The remaining events were classified as {\it M=Medium Energy}.
These cuts were optimized using two large samples of real atmospheric
muons: $i)$ downward throughgoing (average energy of
$\sim$ $300\, \hbox{\rm GeV}$ \cite{trdmacro}), and $ii)$ downward going stopping
muons (average energy of $\sim$$1\, \hbox{\rm GeV}$). 

The aforementioned cuts were applied on the 
upward throughgoing muons (crossing the whole apparatus, {\it lower detector} +
{\it Attico}), both real and simulated. They were selected from the same analysis chain \cite{macroneutrino} and had the same data format.
From the simulation we expect 430 events:  
178 of {\it Low} energy, with $<$$E_\nu$$>=$$11\, \hbox{\rm GeV}$, 
59 of  {\it Medium} energy, with $<$$E_\nu$$>=$$33\, \hbox{\rm GeV}$ and
193 of {\it High} energy, with $<$$E_\nu$$>=$$72\, \hbox{\rm GeV}$.
From the real data, 316 events were selected:
101 of {\it Low}, 51 of {\it Medium} and 164 of {\it High} energy.

The $L$ and $H$ events were further divided according to their zenith angle
$\Theta$: events from the vertical direction $-1.<\cos\Theta<-0.8$, and events with $\cos\Theta>-0.8$. 
For each event topology the number of detected and expected events were determined and ordered with decreasing value of the $<$$L_\nu$$>$/  $<$$E_\nu$$>$ ($<$$L_\nu$$>$= average value of neutrino path length).
The relative systematic uncertainty on each of the five  $<$$L_\nu$$>$/$<$$E_\nu$$>$ values is 12.5$\%$, coming from the uncertainties on the neutrino spectrum and angular shape (discussed in Sect. 4.3), from the detector related effects and analysis cuts. This includes the number of planes used for the refit, the definition of the cuts, the fluctuations in the streamer tube and scintillator efficiencies, and detector acceptance uncertainties. 

We evaluated $\chi^2$ for the hypothesis of no neutrino oscillations using the five $<$$L_\nu$$>$/$<$$E_\nu$$>$  bins, plus the additional point from the analysis of semi contained upward going
events (IU)~\cite{lownu} (upgoing neutrinos with $<$$E_{\nu}$$>$$\sim$$4\, \hbox{\rm GeV}$). 
We found that the distribution agrees with the no oscillation hypothesis
with a probability lower than 2\%). 
For two-flavor oscillations with parameters $\Delta {\rm m^2} = 2.5 \times 10^{-3}$ {\rm e$V^2$} and $\sin^2 2\theta = 1$ we get a probability of 45\%.
\section{The second analysis: streamer tubes in drift mode}

The performance of the streamer tube system, read out in drift mode,
is described in~\cite{macronim}. Here
an absolute muon energy calibration was 
performed at the PS-T9 and SPS-X7 beams, where a slice of the MACRO
detector was reproduced. 
The MCS information was handled 
by a neural network, based on JETNET 3.0\cite{jetnet}, calibrated with the muon beams quoted above.
The neural network(NN) output obtained using test beam data, was compared 
with that expected from
the Monte Carlo simulation, obtaining a satisfactory agreement\cite{macronim}.
The application of such analysis to the MACRO data resulted
in an improvement of the space resolution from $\sigma_{x}$$\simeq$$1\, \hbox{\rm cm}$ 
to $\sigma_{x}$$\simeq$$ 0.3\, \hbox{\rm cm}$.

\begin{figure}[t]
 \begin{center}
  \vskip 3.1cm
  \mbox{\epsfig{file=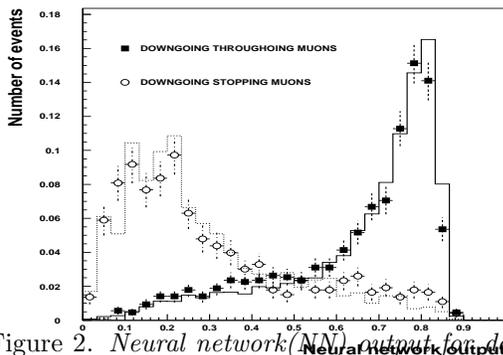,width=7.5cm,height=5.5cm}}
  \vskip -3.1cm
  \caption{\it Neural network(NN) output for down-throughgoing muons:
Real data~(black squares) and 
  Monte Carlo expectations (continuous line).
NN output for downward going
stopping muons: 
real data(empty circle) and
Monte Carlo expectations(dotted line)\label{fig:confronto}.}
\vspace{-0.7cm}
 \end{center}
\end{figure}

Figure~\ref{fig:confronto} shows the Monte Carlo
prediction for NN downward  
throughgoing muons and downward going
stopping muons compared with experimental data. 
A nice agreement between simulation and real data is found for both
categories, 
representing a large energy interval.
 
The result of the neural network output energy calibration
shows that NN output is almost linear with $\log_{10}(E_{\mu})$, 
increasing with the muon energy
up to $E_{\mu}$$\simeq$$40\, \hbox{\rm GeV}$, where a saturation effect occurs. 
With respect to the approach used in\cite{macronim} few details
of the neural network were optimized: the neural network training
and the energy calibration was performed separately for events
with hits in the upper part of the detector({\it Attico}).

\begin{figure}[t]
\begin{center}
\vskip 3.1cm
\mbox{\epsfig{file=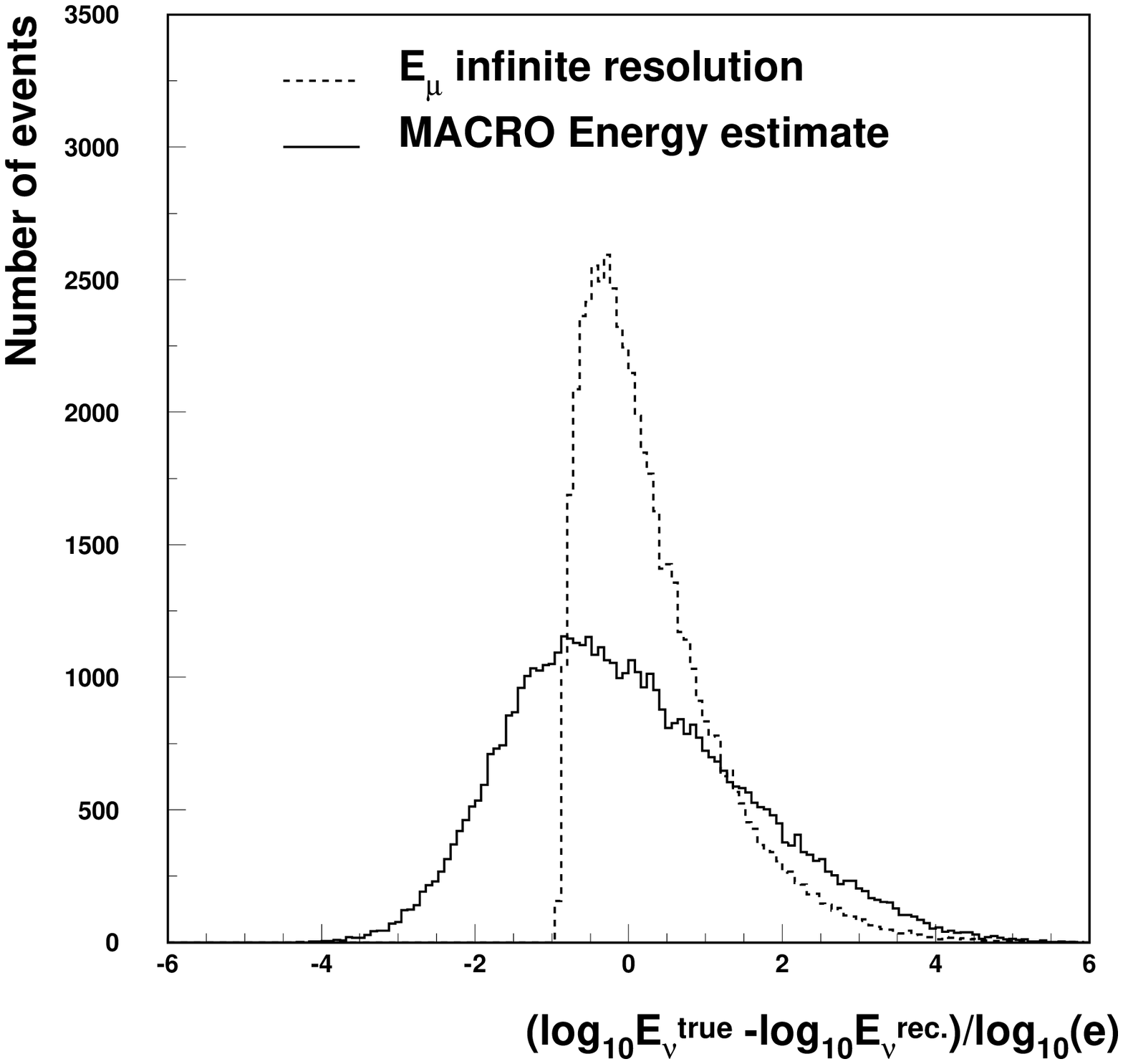,width=7.5cm,height=5.5cm}}
\vskip -3.1cm
\caption{\it Neutrino energy resolution that can be obtained with an 
ideal residual muon energy resolution (dotted line) and with the MACRO
energy estimate(continuous line).\label{fig:stimarisoluzione}}
\vspace{-0.7cm}
\end{center}
\end{figure}

Although smeared by the energy carried away by hadrons and
by energy loss in the rock, the detected neutrino induced 
muons still carry on average $\simeq$40$\%$ of the original 
neutrino energy.
By using the full Monte Carlo simulation quoted in Sect. 2, we calibrated
the NN output
as a function of $\log_{10}(E_{\nu}$): the calibrated NN output is linear
up to $\log_{10}(E_{\nu}(\rm GeV))$$\simeq$2.15. 
We may write:
\begin{equation}
\delta~\log_{10}(E_{\nu})=\log_{10}(e)~\delta \ln {E_{\nu}}=\log_{10}(e)\frac{\delta E_{\nu}}{E_{\nu}},
\end{equation}
where $\delta\log_{10}(E_{\nu})$ is the difference between the ${\rm log_{10}}$
of the real and of the reconstructed neutrino energy.
Taking into account that the NN output was calibrated as a function
of $\log_{10}(E_{\nu})$,
the energy resolution ${\delta E_{\nu}}$/${E_{\nu}}$ was obtained by
plotting the quantity ${\delta~\log_{10}(E_{\nu})}$/${\log_{10}(e)}$.
In Fig.~\ref{fig:stimarisoluzione} we show the resolution that could be
obtained with an ideal muon energy resolution(dotted line)
and that obtained with the present analysis(continuous line). 
The precision of the neutrino energy estimate obtained with 
an ideal muon energy resolution detector is
$\delta E_{\nu}$$/$$E_{\nu}$$\simeq$70$\%$, while with 
the present method 
a resolution of $\delta E_{\nu}$/$E_{\nu}$$\simeq$150$\%$ is obtained. The
asymmetry present in both curves comes from neutrino interactions occurring far from the
detector, for which a large fraction of the muon energy is lost during
the transport.

\subsection{Data Selection}
We used the 783 upward throughgoing muon data set collected in the full 
detector run started
in 1994. 
Table 1 describes further event selection to arrive at the sample for MCS analysis.
We required a single track in both the wire and the
strip view. We selected hits belonging to the track 
and made of a single fired tube, 
to associate unambiguously its QTP-TDC time information. 
This cut is effective
for muon tracks with large zenith angles ($\Theta$ $>$ 30$^\circ$), while it
is quite loose around the vertical: we restricted the present analysis to
events with $\Theta$$\leq$$60^\circ$. Comparisons were performed
between simulated and experimental downward going muon data, to ensure that
the selection efficiency, in the used angular window, is the same in the 
Monte Carlo and in the real data. 
Spurious background hits have been avoided by requiring a time window 
of 2 $\mu$s around the trigger time.
Finally, we selected events with at least four streamer tube planes with 
valid QTP-TDC data. 
We fitted the drift circles using the same tracking developed to analyse test
beam muons. A minimum path length of $200\, \hbox{\rm cm}$ in the {\it lower detector} is 
required for 
tracks hitting the $Attico$ and $400\, \hbox{\rm cm}$ for tracks not hitting it.
These geometric requirement ensure a minimum depth of material
where muons may experience a measurable amount of multiple scattering 
and a lever arm long enough for a comfortable tracking. 
After the selection cuts 
300 events survived, giving an overall efficiency of 38.3$\%$.
\begin{table}[thb]
\begin{center}
\begin{tabular}{|c|c|}
\hline
Cuts&Number of\\
&events\\
\hline
Total number & 783\\
of upward &\\
throughgoing muons&\\
\hline
Single track in the& 695\\
wire and strip views&\\
\hline
$\geq$4 planes with&347\\
valid TDC hits &\\
\hline
Track length cut&314\\
\hline
$\Theta$$\leq$$60^\circ$& 300\\
\hline
\end{tabular}
\vskip 0.5cm
\caption{\small Data selection cuts and selection efficiencies.}
\end{center}
\label{tab:tagli}
\end{table} 
\subsection{Qualitative tests}

We used the information provided by the neural network to separate the
neutrino events into 
four energy subsamples, as shown in Table 2.
The same selection was applied to simulated events.
\begin{table}[thb]
\begin{center}
\begin{tabular}{|c|c|c|}
\hline
Sample&Energy cuts &Median energy\\
&(${\rm GeV}$)&(${\rm GeV}$)\\
\hline
Low&$E_{\nu}^{rec}$$<$30&13\\
\hline
Medium-Low&30$<$$E_{\nu}^{rec}$$<$80&36\\
\hline
Medium-High&80$<$$E_{\nu}^{rec}$$<$130&88\\
\hline
High&$E_{\nu}^{rec}$$>$130&146\\
\hline
\end{tabular}
\vskip 0.5cm
\caption{\small Subsamples selected according with the reconstructed
neutrino energy.}
\end{center}
\label{tab:tagli2}
\end{table} 
\begin{figure}[t]
 \begin{center}
\vskip 4.3cm
\mbox{\epsfig{file=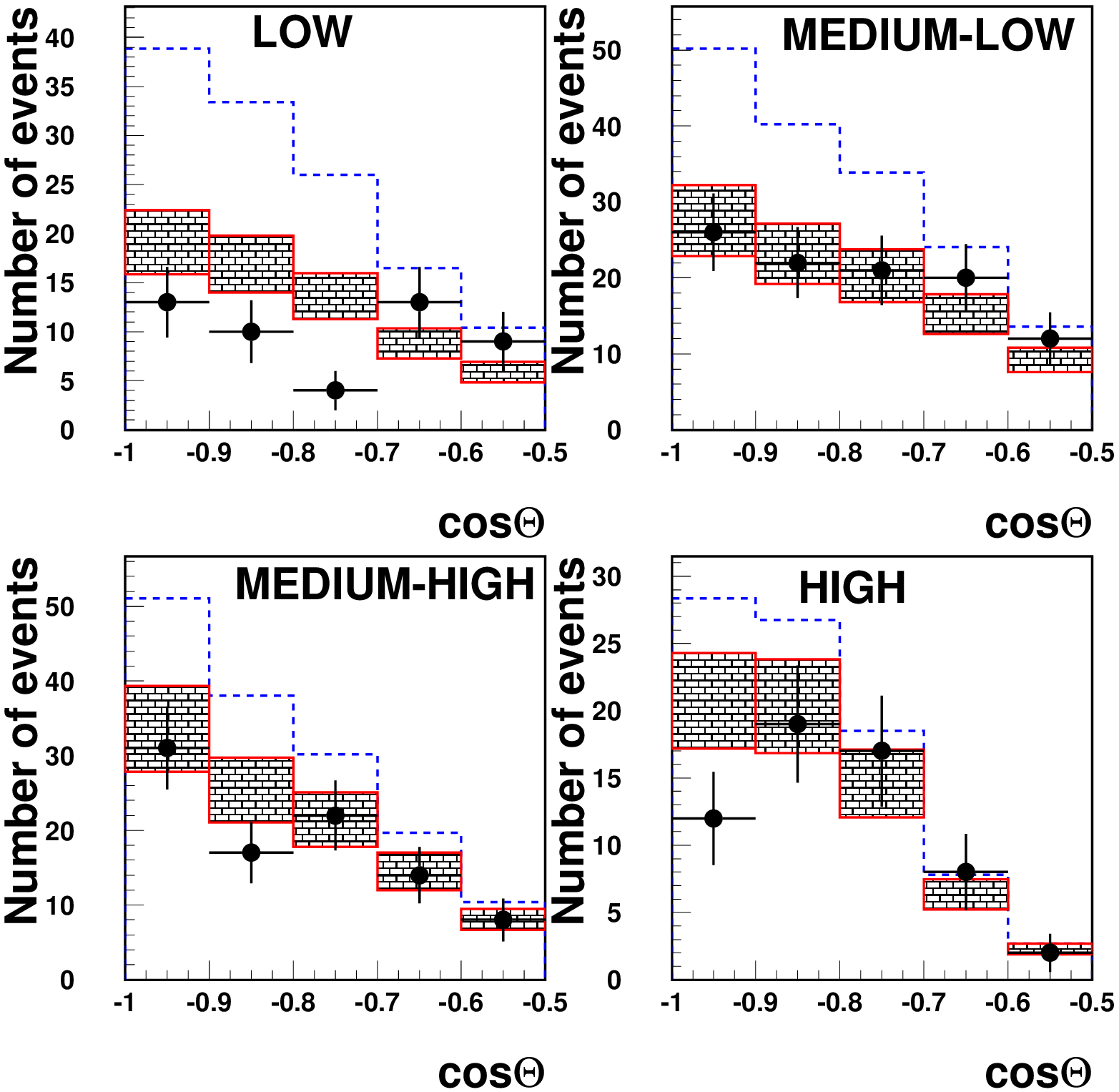,width=9.5cm,height=6.5cm}}
\vskip -3.8 cm
\caption{\it
Number of events versus the cosine of the zenith angle $\Theta$ for
four energy ranges. Black points are the real data, dotted line is the Monte Carlo
simulation, assuming no oscillation, and dotted boxes are the Monte Carlo expectation with
$\Delta {\rm m^2}$=2.5$\times$
$10^{-3}\, \hbox{\rm $eV^2$}$ and $\sin^22\theta=1$,
including a 17$\%$ error.\label{fig:fig1}}
  \vspace{-1.0cm}
 \end{center}
\end{figure}
\begin{figure}[h]
  \begin{center}
    \leavevmode
    \epsfig{file=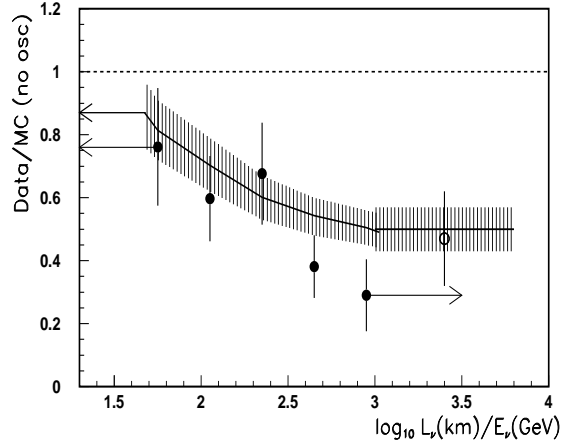,width=6.5cm,height=5.5cm}
      \vskip -1.2cm
\caption{\it Ratio (Data/MC) as a function of the estimated
$L_{\nu}/E_\nu$ for the MACRO upward throughgoing muon sample. The black
circles are the real data over the MC(no oscillation), 
the solid line is the MC assuming
$\Delta {\rm m^2}$~=~2.5$\times$
$10^{-3}$$\, \hbox{\rm $eV^2$}$ and $\sin^22\theta$=1 over
the MC with no oscillation. 
The shaded region represents the MC uncertainties.
The last point(empty circle) is obtained from semicontained
upward going muons.\label{fig:lsue}}
\vskip -1cm
  \end{center}
\end{figure}
Figure 4 shows the zenith angle distributions of the upward throughgoing muons in
the four energy subsamples, compared with the expectations of the Monte Carlo
simulation, assuming no-oscillations(dotted line) and
oscillations with parameters 
$\Delta {\rm m^2}$=
2.5$\times$$10^{-3}$\, \hbox{\rm $eV^2$} 
and $\sin^22\theta\simeq$1
(wall boxes).
The Monte Carlo including the oscillation hypothesis, with the parameters
quoted above, reasonably reproduces the real data in each subsample. 
We point out the strong difference between the oscillations/no-oscillations
hypotheses at low energies, while such difference is reduced 
by increasing the reconstructed neutrino energy.

Finally, we used information on the ratio $L_{\nu}/E_{\nu}$.
The distance $L_{\nu}$, travelled by the neutrinos from production to the
interaction points, 
was measured by MACRO relying on the muon zenith
angle determination, with a precision $\Delta L_{\nu}$/$L_{\nu}\sim~3\%$.
The resolution
on the ratio $L_{\nu}/E_{\nu}$ is therefore
fully dominated by the uncertainty in the neutrino energy estimate, giving
a relative error of $\simeq 150\%$. 
The ratio DATA/Monte Carlo as a function 
of $\log_{10}(L_{\nu}/E_{\nu})$,
is plotted in Fig.~\ref{fig:lsue}. 
The black circles
are the real data over the Monte Carlo predictions, assuming no oscillations, 
the shaded regions represent the Monte Carlo predictions with oscillations,
$\Delta {\rm m^2}$ = 2.5$\times$$10^{-3}$eV$^2$ and $\sin^22\theta$=1, divided by
the Monte Carlo predictions with no oscillation. 

The $\log_{10}$($\frac{L_{\nu}}{E_{\nu}}$) distribution of the neutrinos detected by MACRO spans
from 0.8 to 3.5.
The left-pointing arrow at low $\log_{10}$($\frac{L_{\nu}}{E_{\nu}}$)
represents the effect of the neural network saturation.
Since the maximum energy reconstructed by the NN is $E_{\nu}$$\simeq$$140\, \hbox{\rm GeV}$ 
and since the minimum $L_{\nu}$ for the present
analysis is $L_{\nu}^{min}$$\simeq$6500 km,
the minimum estimated value is
$\log_{10}$($\frac{L_{\nu}}{E_{\nu}})^{min}$$\simeq$1.7.
As far as the right-pointing arrow at high $\frac{L_{\nu}}{E_{\nu}}$ is concerned,
the reconstruction of neutrino energy
based on the residual muon energy, does not allow one
to reconstruct the ratio L/E beyond 
a given limit.
This is due to far neutrino interactions, originated by
high energy muons, which lost a large fraction of its
energy. An ideal detector with ideal muon residual energy resolution, 
requiring $E_{\mu}<$$2\, \hbox{\rm GeV}$, would select 
an average $\log_{10}$($\frac{L_{\nu}}{E_{\nu}}$)=3.2.
Finally, the neural network resolution in muon energy
estimate results in a maximum value $\log_{10}$($\frac{L_{\nu}}{E_{\nu}}$)=3.
The dashed line at 1 is the
expectation without oscillations. The last point(empty circle) is obtained
from semicontained
upward going muon rate\cite{lownu}.
It is not used for the evaluation
of the oscillation probabilities (Sect. 4.4) nor in the allowed region
plot(Fig. 7).
Good agreement is found with the
oscillations expected with the parameters quoted above.

\subsection{Experimental results}
To quantify the contribution of the 
multiple scattering measurement in stand-alone mode, 
we performed a blind analysis using the Monte Carlo data,  
looking for the variable, based
on the energy estimate, showing the maximum sensitivity to separate
the oscillation from the no-oscillation hypothesis.
We found that the best performance is given by the ratio:
\begin{equation}
R=N_{low}/N_{high}
\end{equation} 
where $N_{low}$ and $N_{high}$ are the number
of events with $E_{\nu}^{rec}$$<$$30\, \hbox{\rm GeV}$ and 
$E_{\nu}^{rec}$$>$$130\, \hbox{\rm GeV}$ respectively.

We considered systematic uncertainties due to the neutrino flux calculation
and neutrino cross section.
Due to the large uncertainty in the absolute flux
\cite{Imax,Caprice,AMS,Bess},
we use in this paper only the angular 
distribution and the ratio between different 
event categories selected according to the reconstructed energy.
Nevertheless, the theoretical uncertainty of the existing
neutrino flux, based on the old CR spectrum\cite{Bartol}, has to be accounted for.
We varied the input primary cosmic ray spectral index $\gamma$ in our
simulation by $\Delta \gamma$=$\pm$0.05,
obtaining a theoretical error on R, 
$\left(\frac{\Delta R}{R}\right)_{flux}$=$\pm$13$\%$.

Another source of systematics comes from the neutrino cross section.
We looked at R varying
the Monte Carlo input cross sections. The most important contribution to the
error comes from the ``Low'' category, since the cross section at low
neutrino energy is more uncertain. 
By comparing the cross section computed under several hypotheses, varying
for instance 
the structure function (\cite{GRV94},\cite{Morfin}) in the deep inelastic
scattering, including or neglecting the contribution of resonant
scattering, we found 
$\left(\frac{\Delta R}{R}\right)_{\sigma}$=9~$\%$. 
We estimated a total theoretical error 
$\left(\frac{\Delta R}{R}\right)_{theo}$=
$\sqrt{\left(\frac{\Delta R}{R}\right)^2_{flux}
+\left(\frac{\Delta R}{R}\right)^2_{\sigma}}$=16$\%$.
The systematic error on R reconstruction, evaluated in 6$\%$,
includes the uncertainties on 
absorber density, drift velocity, streamer tube efficiency 
and detector acceptance.
\begin{figure}[h]
 \begin{center}
\vskip 1.4 cm
  \mbox{\epsfig{file=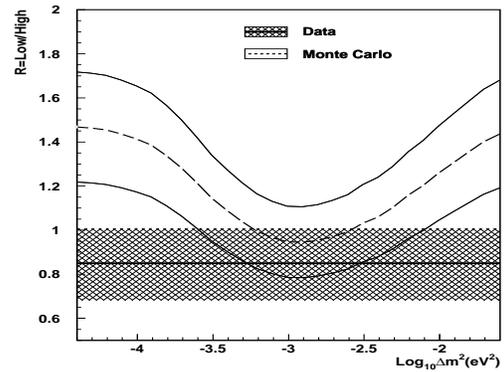,width=7.5cm,height=5.5cm}}
\vskip -1.1cm
  \caption{\it
Low energy events over high energy events as a function of $\Delta
{\rm m^2}$: the area between the solid lines 
includes a 17$\%$ systematics(the error is
not gaussian, see text). The hatched band
  represents the real data.}
\label{fi:lowhigh}
  \vspace{-0.7cm}
 \end{center}
\end{figure}   

Figure~\ref{fi:lowhigh} shows the ratio R as a function of $\Delta {\rm m^2}$, assuming
maximal mixing, for the Monte Carlo simulation: the area
between the two solid curves represents the 17$\%$ systematic error. 
The Monte Carlo prediction in
case of no oscillations is $R^\prime $=1.5$\pm$0.25(theo.+sys.), while for 
$\Delta {\rm m^2}$=2.5$\times 10^{-3}eV^{2}$ and $\sin^{2}2\theta$ =1,
R=1.00$\pm$0.17(theo.+sys.). 
As pointed out in\cite{sterile}, the ratio does not have a Gaussian
distribution.
The errors on the ratio are therefore also not Gaussian: they are reported just to give
a crude estimates of the significance.
The experimental ratio is
$R^{exp}$=0.85$\pm$0.16(stat.).
The one sided probability of measuring a value smaller than the measured
one, was computed by using two different methods. In the first one
we let the Monte Carlo simulation~(no oscillations) to fluctuate
according to statistical and systematic errors of the 
considered ratio. We 
then evaluated the fraction of events giving a value smaller
than the measured one. In a more pessimistic view, rejecting the 
hypothesis that the lower number of events detected by MACRO
with respect to the Monte Carlo expectation
is due to oscillation effects, 
the total number of Monte Carlo events (sum of the numerator and 
the denominator in the ratios), 
were normalized to the experimentally measured sum and then
they were allowed to fluctuate. 

The corresponding one sided probability of measuring a value smaller 
than $R^{exp}$ according to method 1 (2) 
is 0.75$\%$(1.9$\%$) corresponding to 2.4$\sigma$(2.1 $\sigma$).
\begin{table*}[thb]
\begin{center}
\begin{tabular}{|c|c|c|c|}
\hline
&&&\\
Ratio&$R'=R^{MC}_{osc}$&$R''=R^{MC}_{No~osc}$&$R^{exp}$\\
\hline
&&&\\
$\frac{N_{low}}{N_{high}}$&1.00$\pm$0.17&$1.50\pm0.25$&
0.85$\pm$0.16\\
&&&\\
\hline
&&&\\
$\frac{N_{vert}}{N_{hor}}$&1.70$\pm$0.14&2.20$\pm$0.17&
1.48$\pm$0.13 \\
&&&\\
\hline
\end{tabular}
\caption{\small Real Data and Monte Carlo ratios R=$N_{low}/N_{high}$ and
A=$N_{vert}/N_{hor}$.}
\end{center}
\label{tab:digi2}
\end{table*} 
Finally, we combined the information
coming from the energy estimate and from the angular distribution.
We considered the ratio:
\begin{equation}
A=N_{vert}/N_{hor},
\end{equation} 
where $N_{vert}$ is the number of upward throughgoing muon events with $\cos\theta$$\leq$-0.7 and
 $N_{hor}$ is the number of events with $\cos\theta$$\geq$-0.4, as 
discussed in \cite{sterile}.  
The probability of measuring a
value $A^\prime $ lower than $A^{exp}$ according to the method 1~(2) is
P(A'$<$$A^{exp}$)=0.001$\%$(0.01$\%$),
corresponding to 4.3~$\sigma$(3.7$\sigma$).
It is worth noting that, combining the two independent probabilities on the
ratios R and A, the 
probability that a fluctuation of the expected values(no oscillations)
generates the experimental results is 
P($A^\prime$$\cdot$$R^\prime$$<$$A^{exp}$$\cdot$$R^{exp}$)=1.3$\times$$10^{-6}$(3.2$\times$$10^{-5}$),
corresponding to 4.7~$\sigma$(4 $\sigma$).

\begin{figure}[h]
 \begin{center}
 \vskip 4.1cm
\mbox{\epsfig{file=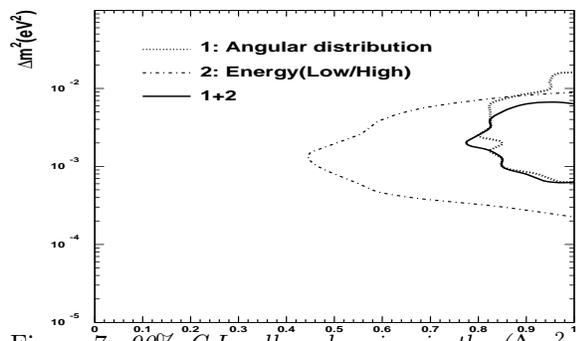,width=8.5cm,height=5.5cm}}
\vskip -3.1cm
  \caption{\it 90$\%$ C.L. allowed region in the ($\Delta {\rm m^2}$, si$n^{2}2\theta$) plane
for $\nu_{\mu}$$\rightarrow$$\nu_{\tau}$ oscillations,
obtained with different data samples\label{fi:esclusione}.}
  \vspace{-0.5cm}
 \end{center}
\end{figure}

The 90\% confidence level allowed regions in the oscillation parameter space have been 
computed according to the prescription
given in \cite{feldman}. 
Figure~\ref{fi:esclusione} shows the 90$\%$ C.L. for the 
ratio R, the angular distribution~\cite{sterile} and for their combination.
\section{Conclusions}
Muon multiple coulomb scattering has been used to estimate the energy of the
neutrino induced upward throughgoing muons detected by MACRO. The different
tests performed on the data, using the estimated energy (separation in
subsamples with different energies, $L_{\nu}/E_{\nu}$ estimates, etc.) give a consistent picture,
all of them supporting the neutrino oscillation hypothesis with parameters
$\Delta {\rm m^2}=$ 2.5$\times$$10^{-3} eV^{2}$ and $\sin^22\theta$$\simeq$1.
To quantify such effect, the ratio $R~=~N_{low}/N_{high}$ was used, in
stand-alone mode and in combination
with the angular distribution, $A=N_{vert}/N_{hor}$. 
Both of them are independent of the neutrino absolute flux. 
The 
significance of the MACRO observation of the neutrino oscillations 
is above 4$\sigma$. 
\vskip 1.5cm
{\bf Acknowledgements}

We gratefully acknowledge the support of the Director and of the staff of the Laboratori Nazionali del Gran Sasso and the invaluable assistance of the technical staff of the Institutions participating in the experiment. We thank the Istituto Nazionale di 
Fisica Nucleare (INFN), the U.S. Department of Energy and the U.S. National Science Foundation for their generous support of the MACRO experiment. We thank INFN, ICTP (Trieste), WorldLab and NATO for providing fellowships and grants (FAI) for non Italian 
citizens.

\end{document}